\documentclass[twocolumn]{aastex631}
\usepackage{amsmath}

\providecommand{\sorthelp}[1]{}

\newcommand{\Planck}{\textit{Planck}}
\newcommand{\Gaia}{\textit{Gaia}}

\newcommand{\sandtheta}{\Theta} 
\newcommand{\thetamax}{\Theta_{\mathrm{max}}}

\newcommand{\lamg}{0.48}
\newcommand{\lamr}{0.62}

\newcommand{\dihea}{\ifmmode \phi \else $\phi$\fi}

\newcommand{\frankie}{FRaNKIE}
\newcommand{\pha}{\ifmmode I_{\mathrm{ph}} \else $I_{\mathrm{ph}}$\fi}
\newcommand{\polpha}{\ifmmode Q_{\mathrm{ph}} \else $Q_{\mathrm{ph}}$\fi}
\newcommand{\nside}{\ifmmode N_{\mathrm{side}} \else $N_{\mathrm{side}}$\fi}

\newcommand{\rinten}{$\mathcal{I}$} 
\newcommand{\gval}{$\mathcal{G}$} 

\received{September 26, 2023}
\accepted{November 1, 2023}
\submitjournal{ApJ}

\shorttitle{Scattered light polarization of Galactic cirrus}
\shortauthors{Bowes and Martin}

\graphicspath{{./}{figures/}}

\begin{document}

\title{Diagnostics from polarization of scattered optical light from Galactic infrared cirrus}

\correspondingauthor{Shannon Bowes}
\email{skbowes@mta.ca}

\author[0009-0008-2598-0873]{Shannon K. Bowes}
\affiliation{Canadian Institute for Theoretical Astrophysics, University of Toronto, 60 St. George Street, Toronto, ON M5S 3H8, Canada}
\affiliation{Department of Physics, Mount Allison University, 67 York Street, Sackville, NB E4L 1E4, Canada}

\author[0000-0002-5236-3896]{Peter G. Martin}
\affiliation{Canadian Institute for Theoretical Astrophysics, University of Toronto, 60 St. George Street, Toronto, ON M5S 3H8, Canada}

\begin{abstract}
We propose polarization of scattered optical light from intermediate Galactic latitude infrared cirrus as a new diagnostic to constrain models of interstellar dust and the anisotropic interstellar radiation field (aISRF). 
For single scattering by a sphere, with Mie scattering phase functions for intensity and polarized intensity for a dust model at a given wavelength (Sloan $r$ and $g$ bands), and 
with models of anisotropic illumination from the entire sky (represented in HEALPix), we develop the formalism for calculating useful summary parameters for an integrated flux nebula (IFN): average of the phase function weighted by the illumination, polarization angle ($\psi$), and polarization fraction ($p$).
To demonstrate the diagnostic discrimination of polarization from scattered light, we report on the effects of different anisotropic illumination models and different dust models on the summary parameters for the Spider IFN.
The summary parameters are also sensitive to the IFN location, as we illustrate using FRaNKIE illumination models.
For assessing the viability of dust and aISRF models, we find that observations of $\psi$ and $p$ of scattered light are indeed powerful new diagnostics to complement joint modeling of the intensity of scattered light (related to the average phase function) and the intensity of thermal dust emission. However, optically thin IFNs that can be modelled using single scattering are faint and $p$ is not large, as it could be with Rayleigh scattering, and so these observations need to be carried out with care and precision. Results for the Draco nebula compared to the Spider illustrate the challenge.

\end{abstract}


\section{Introduction} \label{sec:intro}

Polarization of scattered optical light from intermediate Galactic latitude infrared cirrus is investigated as a new diagnostic to further constrain models of interstellar dust and the anisotropic interstellar radiation field (aISRF). Sandage (1976) identified the nature of these faint optical diffuse cloud structures, which are called integrated flux nebulae (IFN) because they are illuminated not by the light of single star but by the ISRF of the Galaxy. \citet{zhang} demonstrated the benefits of joint modeling of the intensity of scattered light and thermal emission from dust, focusing for illustration on an optically thin IFN and infrared cirrus called the Spider. 

However, for the available data the models were not perfect and revealed some degeneracy between effects of the properties of the dust and of the aIRSF. Hence our exploration of the potential benefits of additional independent constraints from complementary observations, namely the polarization of light scattered by the IFN.  

In Section \ref{sec:pipeline}, we describe how to calculate the intensity and polarization of light scattered by a sphere resulting from illumination from a single incident direction, introducing the scattering geometry, Stokes parameters, and the phase functions for intensity and polarized intensity, illustrated for various dust models.

We present some models of the aISRF in Section \ref{sec:aniisrf} and describe how the illumination of the IFN from over the entire sky is used to calculate summary parameters for the intensity and polarization of scattered light seen by an observer at the Sun (Earth). 

In Section \ref{sec:resultsspider} we provide illustrative results for an IFN at a fixed Galactic location,\footnote{In particular, the Spider, located at Galactic coordinates ($\ell, b) \sim (135^\circ, 40^\circ$) and distance $d$ approximately 320 pc \citep{marchmartin2023}, thus about height $z = 0.2$ kpc above the Galactic plane.} varying the illumination model for a fixed dust model, and vice versa. 

The summary intensity and polarization parameters also depend on the Galactic location, as demonstrated in Section \ref{sec:varycirrus}.

We end with a discussion, summary, and conclusions in Section \ref{sec:discussion}.

\section{Single scattering by a sphere} \label{sec:pipeline}

In this paper we concentrate on optically thin IFNs, because the 3D distribution of dust within infrared cirrus structures is not known. Lacking knowledge of the shape and alignment of dust in IFNs, a further simplifying assumption is that the dust particles are spherical. 

\subsection{Scattered light} \label{subsec:scatteringangle}

\subsubsection{Geometry} \label{sec:geo}

\begin{figure}[h]
    \centering
    \includegraphics[width=0.3 \textwidth, angle=2.2]{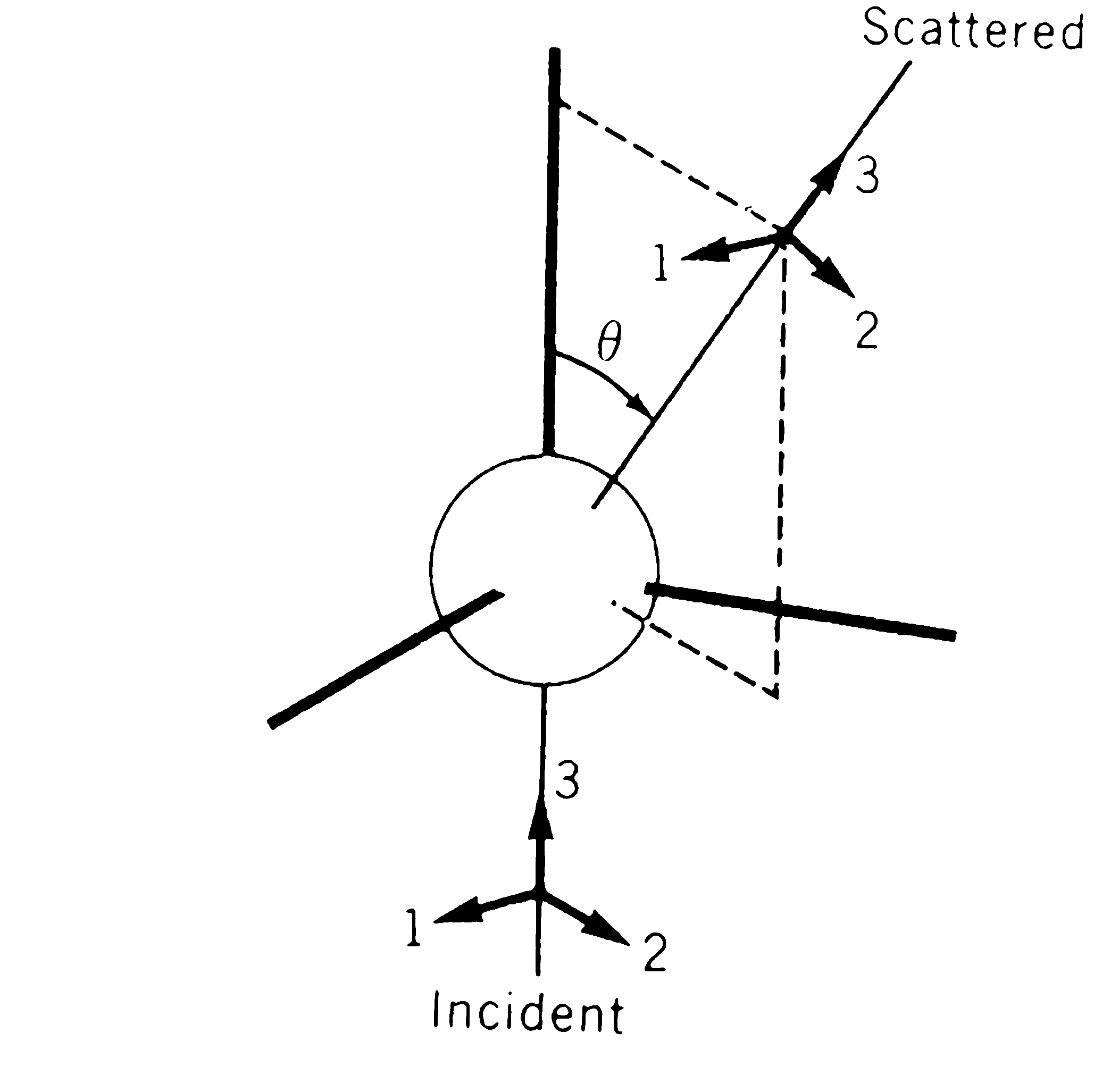}
    \caption{The scattering geometry, including the scattering angle $\theta$, the scattering plane outlined in dashed lines, the two principal orientations of the electric vector, 1 and 2, and the direction of propagation, 3.}
    \label{fig:scatteringangle}
\end{figure}

The scattering geometry is described in Figure \ref{fig:scatteringangle} taken from Figure 4.1 right in \citet{martin78}. The scattering angle $\theta$ is the angle between the forward direction of incoming light (incident interstellar radiation field) and of the outgoing scattered light. The scattering plane is defined to contain these two vectors.  There are two principal orientations of the electric vector, perpendicular and parallel to the scattering plane and denoted 1 and 2, respectively.  The incident intensity is assumed to be unpolarized, and so components 1 and 2 are equal.  However, upon scattering they are not equal and by symmetry the light has become linearly polarized in an orientation either perpendicular or parallel to the scattering plane.

\subsubsection{Stokes parameters} \label{sec:sto}

\begin{figure}
    \centering
    \includegraphics[scale=0.35]{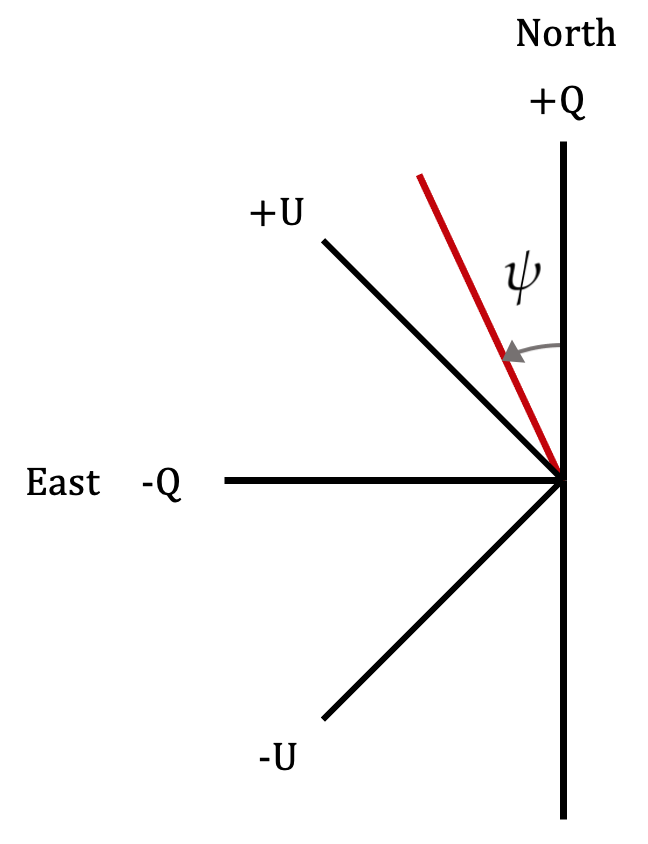}
    \caption{Relationship of the Stokes parameters $Q$ and $U$ for linear polarization to the orientation $\psi$ of the electric vector, measured from North through East. 
    For electric vector orientation along the vertical axis (North-South; $\psi = 0)$, $Q$ is positive and $U=0$, for which we use $+Q$ as shorthand.  
    For electric vector rotated counterclockwise to $\psi = 45^\circ$, we have correspondingly $+U$ ($Q = 0$).
    Negative and positive of each Stokes parameter correspond to electric vector orientations separated by $90^{\circ}$.  At $180^\circ$ the orientation is the same as for $0^\circ$, and so $Q$ is positive and $U=0$.
    The actual polarization, e.g., the red line, is a pseudovector characterized by a length $P$ and orientation (not direction) $\psi$ defined in the half plane, with $\psi$ ranging over [$0^\circ$, $180^\circ$] (see Equation \ref{eq:fracpol}).  The signs of $Q$ and $U$ in each $45^\circ$ sector are indicated by the signs on the boundary black lines, stemming from the ``projections'' $Q = P \cos(2\psi)$ and $U = P \sin(2\psi)$.
    For an observer of scattered light originating as unpolarized light single-scattered by a sphere, ``North" is taken to be along 2 in the scattering plane (Figure \ref{fig:scatteringangle}), and the scattered light is described entirely by $I$ and $Q$ in that frame of reference ($U = 0$).
    }   
    \label{fig:stokes}
\end{figure}

The linear polarization of light can be described using the Stokes parameters, $I$, $Q$, and $U$.
The total intensity is described by $I$ and the linearly polarized intensity by the two parameters $Q$ and $U$.\footnote{Circular polarization is described by $V$; in our application, no circular polarization is produced and so this parameter is not relevant.}   As described more fully in Figure \ref{fig:stokes}, the polarization (red line) corresponding to the parameter pair ($Q$, $U$) is a pseudovector with orientation defined in a half plane, range [$0^\circ, 180^{\circ}$]. For an observer of the scattered radiation,``North" is taken to be along 2 in the scattering plane.

\subsubsection{Polarization of scattered light} \label{sec:pol}

Because the incident light is unpolarized, only Stokes parameter $I_\mathrm{i}$ is non-zero. On scattering by a spherical dust grain, the radiation acquires polarization, with electric vector either parallel or perpendicular to the scattering plane (sp) and so for the scattering plane reference system $U_\mathrm{sp} =0$. The intensities of the incoming light (the illumination) and scattered light are described by the Stokes parameters ($I_\mathrm{i}$, 0, 0) and ($I_\mathrm{s}$, $Q_\mathrm{sp}$, 0), respectively. The transformation maxtrix that acts on the incident ($I_\mathrm{i}$, 0) column vector to polarize the scattered light can be written \citep{martin78}:

\begin{equation} \label{eq:matrix}
F = \frac{C_\mathrm{s}}{4\pi} \times \begin{bmatrix}
\frac{1}{2}(M_2 + M_1) & \frac{1}{2}(M_2 - M_1)\\
\frac{1}{2}(M_2 - M_1) & \frac{1}{2}(M_2 + M_1)
\end{bmatrix}\, ,
\end{equation}

where $C_\mathrm{s}$ is the scattering cross section and $M_1$ and $M_2$ are calculated for incident electric vector perpendicular and parallel to the scattering plane, respectively.  Only the first column of the matrix is relevant because $Q_\mathrm{i} = 0$. In the calculations below, both the normalization factor of the illumination and the factor $C_\mathrm{s}/4\pi$ are ignored, because they affect $I_\mathrm{s}$ and $Q_\mathrm{sp}$ of the scattered light equally. \citet{zhang} discuss how the product of these factors can be related to the brightness of the thermal dust emission.

A familiarizing case re the matrix elements is Rayleigh scattering, where the size of the sphere is much smaller than the wavelength.\footnote{The case where the size of the sphere is much larger than the wavelength is much more complicated and less intuitive, involving diffraction, transmission, reflection, and interference, and is best handled by a good Mie code able to capture these combined effects.} In that case $M_1/2 = 3/4$ and $M_2/2 = 3/4 \times cos^2\theta$.

For our applications, $M_1$ and $M_2$ are calculated using Mie theory from closed-form infinite series that are functions of the scattering angle $\theta$. The dust models have size distributions for a given composition and up to two compositions.  Elements $M_1$ and $M_2$ depend on size and composition, and their combined effect is accomplished through weighting by corresponding cross section $C_\mathrm{s}$, which depends on size and composition as well.

Note that the scattered light distribution depends only on $\theta$ and is axisymmetric around the direction of the incident radiation.  The Mie calculations are carried out on a grid of $\theta$ within which values of the matrix elements at any $\theta$ can be interpolated as needed.  The grid used is every degree from $0^\circ$ to $180^\circ$ inclusive (181 points).

The scattering angle used for the interpolants below was calculated as the $\arccos$ of the dot product on the unit sphere between the incident and scattered light vectors, where in our applications the scattered light of interest is that scattered toward Earth.  We used cubic spline interpolation and comment on this choice below in Section \ref{sec:wholesky}.

\subsubsection{Phase functions for intensity and polarized intensity} \label{sec:phas}

The phase function for intensity \pha\ describes the intensity distribution as a function of $\theta$,  relative to the average of the scattered light intensity integrated over the sphere (with this normalization, the integral of \pha\ over the sphere is 4$\pi$).  In terms of the above,

\begin{equation}
    \pha = \frac{1}{2}(M_2 + M_1)\, .
    \label{eq:pha}
\end{equation}

There is a corresponding phase function for polarized intensity, 

\begin{equation}
    \polpha = \frac{1}{2}(M_2 - M_1)\, 
    \label{eq:polpha}
\end{equation}

in the specified reference frame (\pha\ does not depend on the reference frame).  Both $I_\mathrm{s} = I_\mathrm{i}\,\pha$ and $Q_\mathrm{sp} = I_\mathrm{i}\,\polpha$ scale in the same way with $I_\mathrm{i}$.

For Rayleigh scattering, $\pha = 3/4 \times (cos^2\theta + 1)$ and $\polpha = 3/4 \times (cos^2\theta - 1)$.  Note that in our convention (Figures \ref{fig:scatteringangle} and \ref{fig:stokes}), this \polpha\ is negative, with the electric vector oriented perpendicular to the scattering plane.

\subsection{Phase functions for representative dust models} \label{subsec:dustphasefunctions}

\begin{figure}
    \centering
    \includegraphics[scale = 0.5]{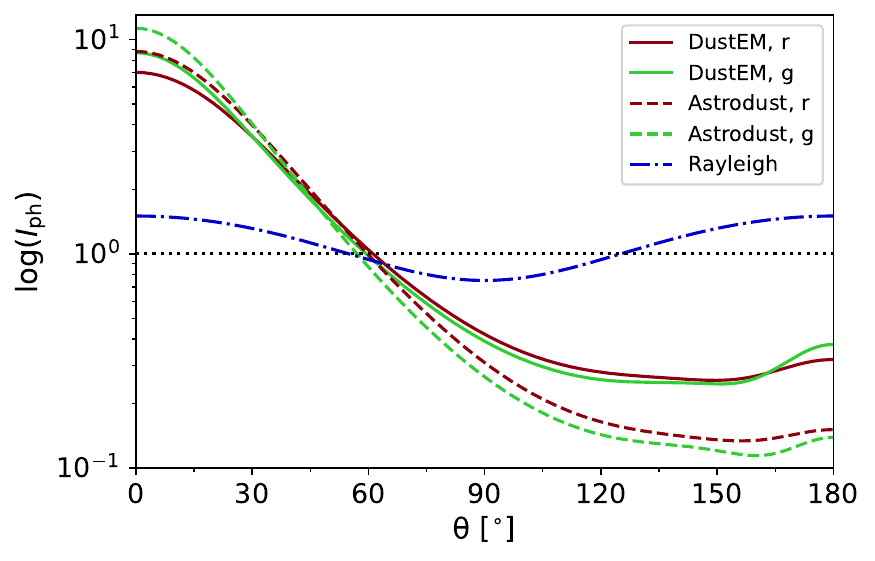}
    \includegraphics[scale = 0.5]{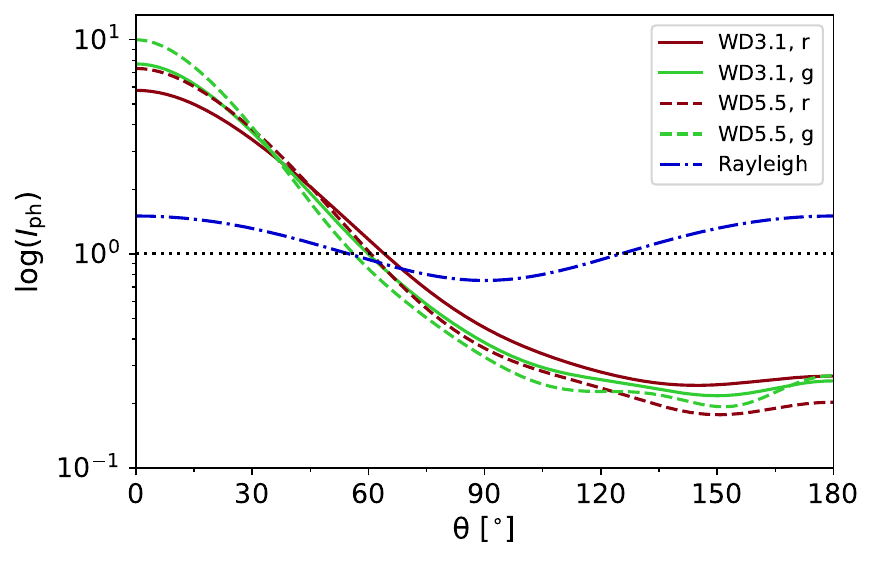}
    \caption{Phase functions \pha\ for various dust models in the $r$ and $g$ bands. The phase functions for Rayleigh scattering and isotropic scattering (unity) are included for comparison.
    Top: DustEM and Astrodust. 
    Bottom: WD3.1 and WD5.5.
    Even in the event of color blindness, the curves in this and following figures are tied to their correct identity through the inset internal labels.
    }
    \label{fig:phasefxns}
\end{figure}

For purposes of illustration, we consider the DustEM dust model \citep{comp11} as used by \citet{zhang} and Astrodust \citep{dh21}. The DustEM model has two separate populations of grains, ``LamC'' and ``aSil'' (large amorphous carbon and amorphous silicate grains), whereas Astrodust has composite grains of amorphous carbon and silicate material; we adopt the Astrodust default model with porosity 0.2 and no iron inclusions \citep{Hensley_2023}.  Composite grains such as these have been discussed before \citep{mathis96,dwek97}. We also consider two models by \citet{WD01} (WD) that like DustEM are composed of large amorphous carbon and amorphous silicate grains. These are historically viable dust models constrained by many observational considerations (e.g., \citealt{hd21}), for which we are seeking further discrimination including their response to different aISRFs (Section \ref{sec:aniisrf}).  

For these and similar dust models the adopted size distributions are constrained by observations such as the wavelength dependence of interstellar extinction. Most match a ratio of total to selective extinction $R_V \approx 3.1$ characteristic of the diffuse interstellar medium, but one (WD5.5) has size distributions reproducing $R_V \approx 5.5$, which is more characteristic of dense regions not encountered at intermediate and high Galactic latitudes. Results from WD5.5 can be compared to those from WD3.1 to show the differential effects from the changing size distribution (larger on average).

\begin{figure}
    \centering
    \includegraphics[scale = 0.5]{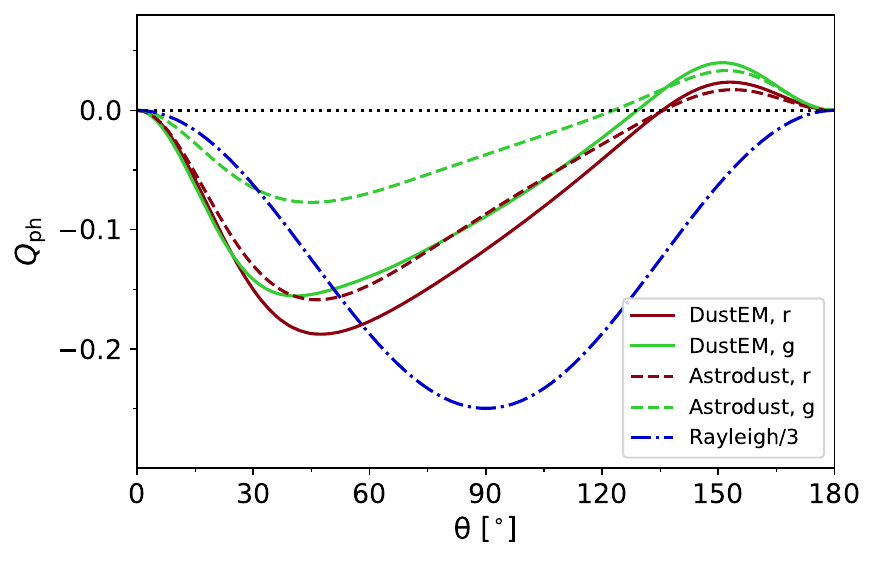}
    \includegraphics[scale = 0.5]{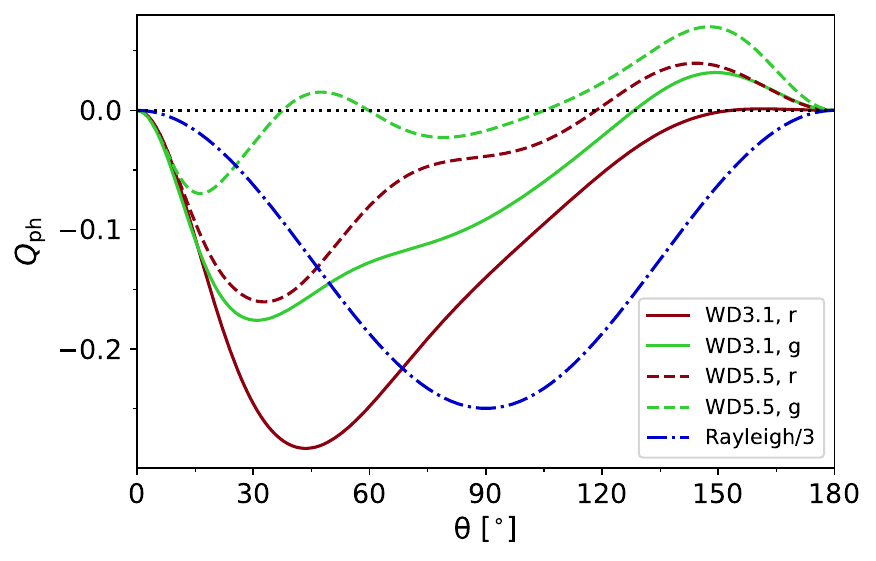}
    \caption{Polarized intensity phase functions \polpha\ for various dust models 
    in the $r$ and $g$ bands. The function for Rayleigh scattering (scaled down by 1/3) is included for comparison.
    Top: DustEM and Astrodust. 
    Bottom: WD3.1 and WD5.5.
    }
   \label{fig:polphasefxns}
\end{figure}

\subsubsection{Phase functions for intensity, \pha} \label{sec:dmpha}

Phase functions \pha\ for these dust models are shown in Figure \ref{fig:phasefxns}, calculated for the Sloan $g$ ($\lambda_0 \sim \lamg$) and  Sloan $r$ ($\lambda_0 \sim \lamr$) bands used by the Dragonfly telescope array \citep{zhang}. These can be contrasted to the phase function for Rayleigh scattering, which is close to isotropic ($\pha = 1$). Forward scattering is heavily favored. Relatively little light is scattered ``sideways'' and there is a slight back-scattering peak.

The strong tendency toward forward scattering can be quantified using the asymmetry parameter, \gval, which is the average $\cos\theta$ weighted by \pha\ over the sphere. This can be evaluated in closed form from Mie theory and averaged over the size distribution.  As seen by the entries in Table \ref{tab:resultsu} in Section \ref{sec:resultsspider} below,\footnote{We tabulate results to many significant figures, not that this high accuracy is important for our conclusions, but to be helpful for anyone reproducing these calculations.  We have computed these by independent pipelines written in IDL and in python.} \gval\ is close to 0.6 in the optical for these dust models, increasing with the ratio of the grain size to the wavelength.  For the Rayleigh phase function \gval\ $ = 0$, just as for isotropic scattering.

We also tabulate the albedo, $\varpi$, which is the ratio of the scattering cross section to the extinction cross section (itself the sum of the scattering cross section and absorption cross section).

\subsubsection{Phase functions for polarized intensity, \polpha} \label{sec:dmpolpha}

Similarly, Figure \ref{fig:polphasefxns} shows the polarized intensity phase functions for the dust models, with the Rayleigh scattering function for comparison.  All functions are zero in the forward and back-scattering directions. Rayleigh scattering, with electric vector perpendicular to the scattering plane (negative \polpha), produces strong polarization in the ``sideways'' direction. Overall, the amplitude for the dust models is much smaller. Peak negative \polpha\ for the dust models occurs in the forward-scattering hemisphere. In the back-scattering hemisphere \polpha\ changes sign (electric vector parallel to the scattering plane) and has a low amplitude positive peak. The behavior is more complex for WD5.5 in the $g$ band, where the ratio of size to wavelength is largest.

\section{Anisotropy of the incident ISRF}
\label{sec:aniisrf}

Even with anisotropic phase functions, the scattered radiation is isotropic and unpolarized unless the incident radiation making up the ISRF is also anisotropic.  For illustrative purposes, we use the same three aISRF models considered by \citet{zhang}.

\begin{figure}
    \centering
    \includegraphics[scale=0.38, trim = 1.1cm 0 0 0]{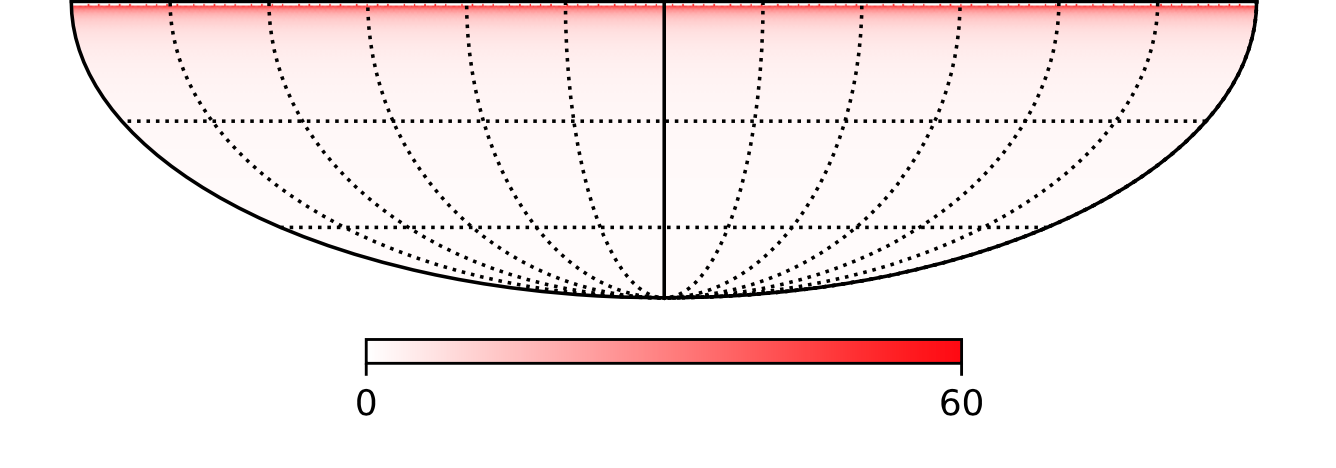}
    \includegraphics[scale=0.38, trim = 1.1cm 0 0 0]{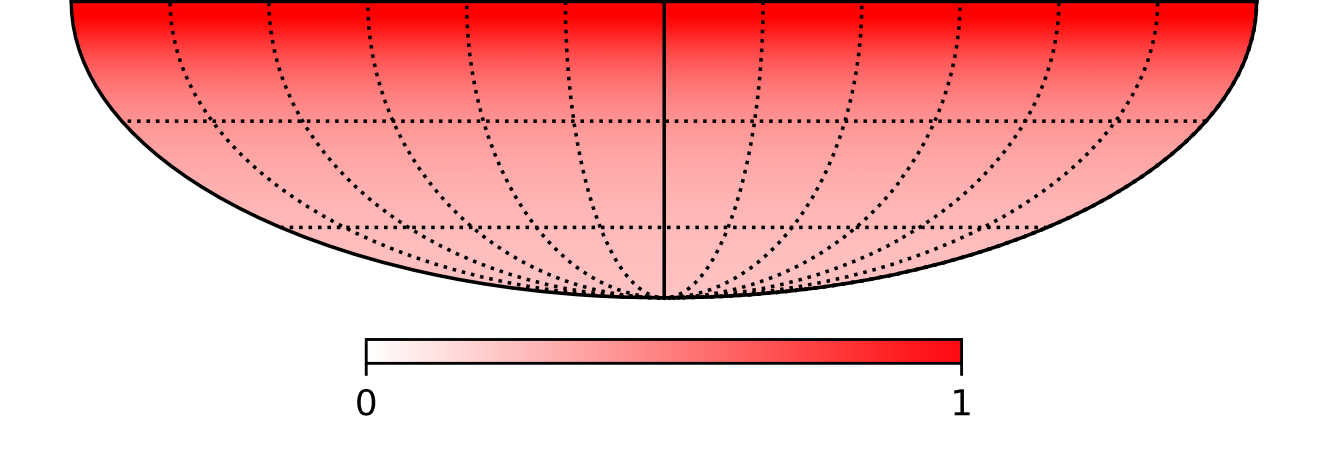}
    \includegraphics[scale=0.38, trim = 1.1cm 0 0 0]{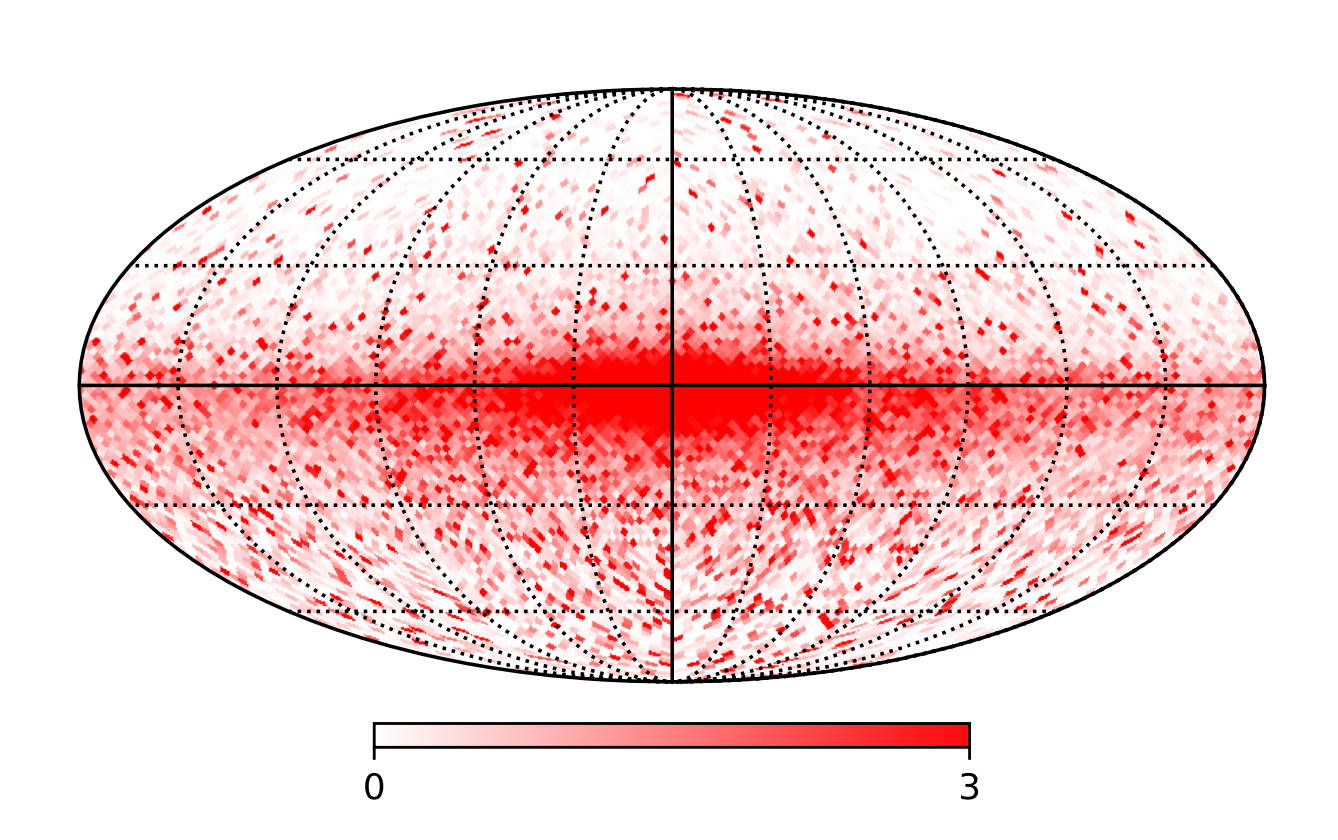}
    \caption{Mollweide projections showing the anisotropy of three different illumination models each using a different relative scale.
    Top: Following \citet{sandage76}, $1/\cos \sandtheta$ with $\sandtheta_{\rm max}= 89^\circ$. 
    Middle: USIE model (relating to \citealt{mezger82} -- see text), $1 - \exp( - 2 \tau_\infty/\cos \sandtheta)$ with $\tau_\infty= 0.139$ in the $r$ band. 
    Bottom: \frankie\ illumination model for Galactocentric distance 8.5 kpc, height 0.2 kpc, and filter at 0.6 \micron\ (near the $r$ band), divided by its average value. 
    }
    \label{fig:sandageillum}
\end{figure}

\subsection{Sandage} \label{sec:sanill}

One illumination model that we tested was proposed in \citet{sandage76}, in which an IFN is illuminated by the entire Galactic plane below, taken to be a slab of uniform brightness and extending to infinity. This produces an axisymmetric illuminating intensity proportional to $1/\cos \sandtheta$, where $\sandtheta$ is measured from the perpendicular dropped from the IFN to the plane. Different asymmetry can be explored by choosing $\sandtheta_{\rm max} < 90^\circ$, thus illumination within a cone from a truncated slab. The total illumination integrated over the lower hemisphere is divergent as $- \log \cos \sandtheta_{\rm max}$ and so Sandage arbitrarily limited $\sandtheta_{\rm max}$ to be 89$^\circ$.  This is the value used in Figure \ref{fig:sandageillum}, top, where over the lower hemisphere the intensity ranges in relative units from unity to about 57.\footnote{Only the relative intensities pixel to pixel matter for the three dimensionless summary intensity and polarization parameters \rinten, $p$, and $\psi$ calculated below (Section \ref{subsec:polcalc}).} To capture such small-scale non-uniformity in a HEALpix representation (Section \ref{sec:wholesky}), an appropriately large $\nside$ is needed.\footnote{The number of pixels is $12 \times \nside^2.$}  A value $\nside = 256$ (786,432 pixels) is sufficient for these models, based on convergence of the computed $p$ and $\psi$ and of the integral of the intensity.

\subsection{Uniform slab with internal extinction (USIE)} \label{sec:uniill}

A second model is like the Sandage model, based on illumination from below by a uniform slab, but with the distinction that it has internal extinction characterized by optical depth $\tau_\infty$ from the midplane vertically to the slab edge. The intensity is then proportional to $1 - \exp( - 2 \tau_\infty/\cos \sandtheta)$.\footnote{Our model uses the same equations as used by \citet{mezger82} to evaluate the intensity seen by an observer at the midplane of a uniform slab, and its analytic integral over a hemisphere, except for the factor two multiplying $\tau_\infty$.} In the limit $\tau_\infty \rightarrow 0$, it becomes the Sandage model for the same $\sandtheta_{\rm max} < 90^\circ$. 

A scale for $\tau_\infty$ is required and we adopted theirs \citep{mezger82, mathis83}, using the CCM extinction curve \citep{cardelli89} for $R_{\textrm V} = 3.1$ to obtain values of $\tau_\infty= 0.139$ and 0.191 for the $r$ and $g$ bands, respectively.  The $r$ filter illumination displayed in Figure \ref{fig:sandageillum}, middle panel, uses $\sandtheta_{\rm max} = 90^\circ$, which can be computed because the internal extinction removes the singularity.  The intensity over the lower hemisphere ranges in relative units from 0.24 to unity, much more uniform (isotropic) than the Sandage model.   Again a value $\nside = 256$ is sufficient.  With increasing $\tau_\infty$ (as in the $g$ band) the illumination becomes even more isotropic. 

\subsection{\frankie} \label{sec:fraill}

Third, we used the suite of aISRF models from the FRaNKIE code, a tool for calculating multi-wavelength interstellar emission in galaxies \citep{porter05,porter15,porter17} as part of the GALPROP code for cosmic-ray transport and diffuse emission production. These \frankie\ models are based on realistic distributions of the sources of the radiation, including in the Galactic bulge, radiative transfer and dust, and the resulting illumination distribution changes with wavelength and location of the IFN (Galactocentric distance and height above the plane). From the grid available on the GALPROP web site site,\footnote{\url{https://galprop.stanford.edu/}, specifically the data file for GALPROP version 54 (and 56), galprop-54\_data\_081715.tar.gz} we downloaded those most appropriate to application to the Spider IFN and the Dragonfly bands: Galactocentric distance 8.5 kpc, height 0.2 kpc, and filters at 0.6 and 0.45 \micron. We also explored results for the illumination at different heights.\footnote{The \frankie\ illumination files used were centered on the Sun (assumed midplane) or at heights directly above. For our calculations, for a model of the appropriate height we performed a simple parallel transport to the location of the IFN because rotation by a small angle (of order a degree) would not qualitatively affect the results.}

The illumination displayed in Figure \ref{fig:sandageillum},  bottom, for the Spider IFN near the $r$ band, is seen to be more realistic than the other two models.  The IFN is illuminated from all directions, not just from below, and there is an azimuthal asymmetry (with Galactic longitude) because the illumination from directions in the inner Galaxy is higher.

\subsection{Rotation to a common reference frame} \label{subsec:rotation}

The polarization pseudovector ($Q_\mathrm{sp},0$) of the scattered radiation is relative to the scattering plane. For a constant scattering direction, toward the Earth, the scattering angle and orientation of the scattering plane change with the direction of incident light. Therefore, when considering more than a single direction of incident light, it is necessary to rotate the pseudovector on the sky about the vector joining the IFN to Earth to a common reference frame in which to sum the contributions to the individual Stokes components of ($Q_\mathrm{s}, U_\mathrm{s}$). We chose the common reference frame to be the meridional plane containing the IFN, the Galactic North pole, and Earth, which we refer to as the sky reference frame with polarization in the Galactic convention. The vector joining the IFN to Earth is a common vector of the two planes, and the required amount of rotation of the column pseudovector is by twice the dihedral angle \dihea\ between the planes.  The rotation matrix, of which only the first column is relevant because $U_\mathrm{sp}=0$ in the scattering plane reference frame, is

\begin{equation} \label{eq:dihea}
\begin{bmatrix}
cos2\dihea \, & -sin2\dihea\\
sin2\dihea \, & \phantom{-}cos2\dihea
\end{bmatrix}\, ,
\end{equation}

where \dihea\ is measured from North through East in the meridional Galactic frame.  The scattered light intensity $I_\mathrm{s}$ is of course independent of the reference frame.

In vector form, the dihedral angle is obtained via the dot product of the unit normals to the two planes and the normal for each plane is obtained from the cross product of the IFN to Earth vector and another vector in that plane. This can be reduced to dot products among the three vectors, appropriately normalized and coded to take into account edge cases: forward and back scattering and coplanar.

\subsection{Using HEALPix over the whole sky} \label{sec:wholesky}

In practice, the whole sky illuminating intensity for \frankie\ is given on a HEALPix sphere \citep{gorski05}, with $\nside = 32$ (12,288 pixels).\footnote{\url{https://healpix.sourceforge.io}} We use HEALPix for the other illustrative aISRFs as well. HEALPix divides a sphere into equal-area pixels. Each indexed pixel corresponds to a single incident light vector as seen by dust in the IFN, with an associated radiation field intensity. We perform calculations of the scattered light intensity and polarized intensity pixel by pixel over the whole sky using the method outlined above, with the scattering direction always toward Earth. Summing the scattered light contributions to $I$, $Q$,  and $U$ in the common reference frame, weighted by the varying incident intensity over the whole sphere, results in a single Stokes parameterization ($I_\mathrm{s,tot},Q_\mathrm{s,tot},U_\mathrm{s,tot}$) describing the scattered light seen at Earth.

For efficiency, the steps in the calculation can be grouped according the goals at hand. Here, the quantities of interest can be produced using vectors on the HEALPix indexed sky. For the scattering geometry, vectors $\vec{\theta}$ and $\vec{\phi}$ and then $\vec{\cos2\phi}$ and $\vec{\sin2\phi}$ can be (pre-)computed for a given $\nside$. Then for different dust models and bands, their vectors $\vec{\pha}$ and $\vec{\polpha}$ can be produced by interpolation to the given $\vec{\theta}$. Different incident aISRFs are represented by their illumination vectors $\vec{I_\mathrm{i}}$. The total scattered light $I_\mathrm{s, tot}$ is obtained via the dot (scalar) product 
$\vec{I_\mathrm{i}} \cdot \vec{\pha}$. The vector describing the polarized intensity in the scattering plane frame of reference $\vec{Q_\mathrm{sp}}$ is obtained by the Hadamard (or Schur or element-wise) product $\vec{I_\mathrm{i}} \circ \vec{\polpha}$ (like the dot product without the summation). The totals $Q_\mathrm{s,tot}$ and $U_\mathrm{s,tot}$ in the common reference frame are obtained by dot products  $\vec{Q_\mathrm{sp}} \cdot \vec{\cos2\phi}$ and $\vec{Q_\mathrm{sp}} \cdot \vec{\sin2\phi}$, respectively. To follow the steps during code development, we examined many intermediate vectors using Mollweide projection.

Simple checks of the accuracy of the interpolation and integration are available.  For example, the numerical integral of the phase function over the sphere (as the sum of the elements of $\vec{\pha}$ divided by the number of pixels, i.e., their average) can be compared to the expectation value unity.  Similarly, \gval\ can be evaluated numerically (as the dot product $\vec{\cos\theta} \cdot \vec{\pha}$ divided by the number of pixels) for comparison to that computed directly from Mie theory. For example, for the typical phase functions used here these numerical integrals show agreement to within one in the fourth figure, using $\nside = 32$ and linear interpolation in a uniform grid based on 181 angles $\theta$.  Cubic spline interpolation yields an order of magnitude improvement even for \gval, and so we adopt that.

\subsection{Summary intensity and polarization parameters} \label{subsec:polcalc}

One useful diagnostic factor is \rinten, a relative intensity obtained by dividing $I_\mathrm{s, tot}$ by the total illumination (the sum of the elements in $\vec{I_\mathrm{i}}$ ).  This is simply the average of the phase function \pha\ weighted by the illumination. The value of \rinten\ is unity for an isotropic radiation field and/or a phase function with \gval\ = 0. A value of \rinten\ $< 1$ means that the observer is in a position less favorable than average relative to the IFN, given the asymmetries of the aISRF and/or \pha, i.e., the IFN is dimmer as discussed by \citet{zhang}. 

Two parameters that describe polarization independent of the scale of the illuminating intensity and the scattering cross section are the polarization fraction $p$, also called the degree of polarization, and the polarization angle $\psi$ (see Figure \ref{fig:stokes}), also called the position angle or orientation of the electric vector. The conventional formulae are

\begin{equation} \label{eq:fracpol}
    p =\frac{\sqrt{Q^2+U^2}}{I}\, \mathrm{and}\, \psi = \frac{1}{2}\arctan(U,Q)\, ,
\end{equation}

where the two-parameter arctangent ensures that for the orientation $\psi$ the correct sector in the half-plane is obtained (Figure \ref{fig:stokes}). We use these formulae on ($I_\mathrm{s,tot}, Q_\mathrm{s,tot}, U_\mathrm{s,tot}$) to obtain the summary parameters reported below.

\section{Results for the Spider IFN} \label{sec:resultsspider}

To demonstrate the diagnostic discrimination of polarization from scattered light, we report on the effects of different illumination and different dust models on the summary intensity and polarization parameters \rinten, $\psi$, and $p$ for the location of the Spider IFN.

\subsection{Varying incident radiation fields}\label{subsec:varyisrf}

With the dust model held fixed, the summary parameters were calculated for the three different aISRFs. For the Sandage and USIE models, because of axisymmetry the results cannot depend on $\ell$ and only the location parameter $b$ is required ($40^\circ$ for Spider). Using a \frankie\ model, both $\ell$ and $b$ need to be specified, as well as the height above the Galactic plane (which implies a distance). These examples illustrate the additional and complementary discriminatory power of polarization observations in assessing proposed aISRFs. 

\begin{figure}
    \centering
    \includegraphics[scale = 0.5]{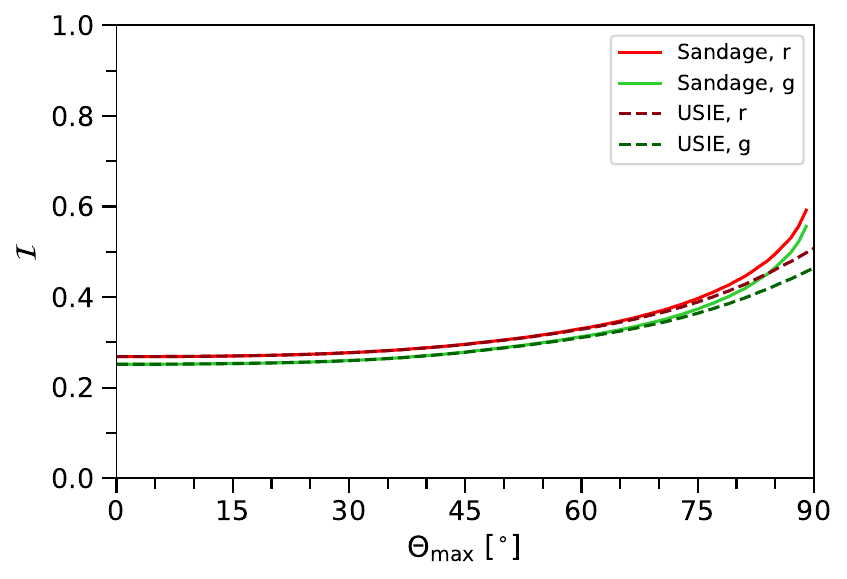}
    \includegraphics[scale = 0.5]{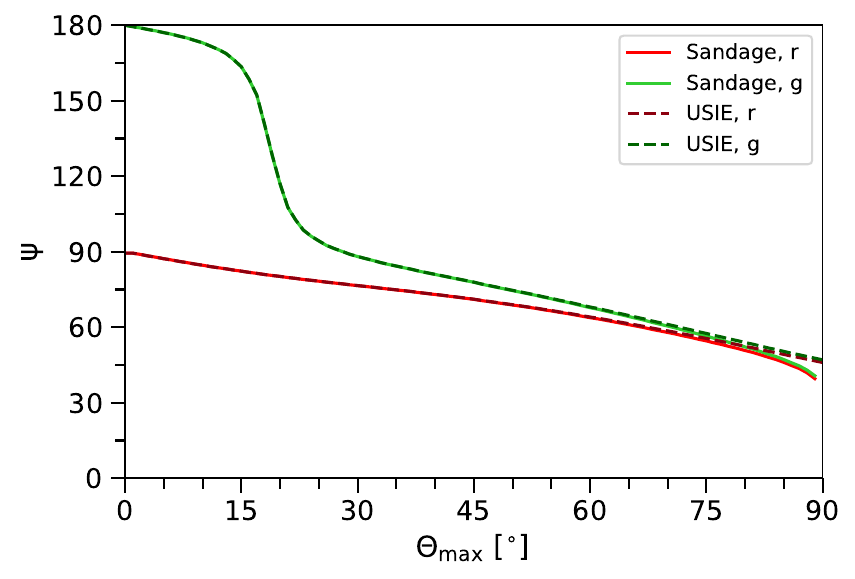}
    \includegraphics[scale = 0.5]{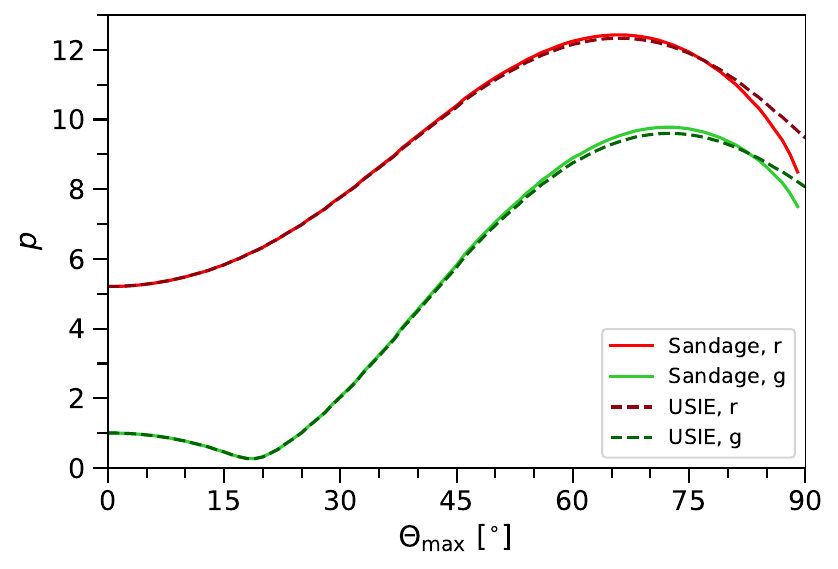}
    \caption{Effect of different illumination models, with internal extinction (USIE) and without (Sandage), on polarization parameters calculated with HEALPix $\nside = 256$ for the DustEM model at $r$ and $g$, and location parameter $b = 40^\circ$. 
    Top: relative intensity. Middle: polarization angle in degrees.
    Bottom: polarization fraction in percent.
    Values for USIE with $\thetamax = 90^\circ$ are tabulated in columns 4 -- 6 and DustEM rows in Table \ref{tab:resultsu}.
    }
    \label{fig:polanglecomparison}
\end{figure}

\subsubsection{Sandage and USIE models of illumination as a function of $\thetamax$ for fixed $b$ }
\label{subsubsec:sandage}

In the Sandage model, the (half) opening angle of the cone within which there is incident radiation is $\thetamax$, and so this can be used to explore different geometries of the radiation field. The dependence of \rinten, $\psi$, and $p$ on $\thetamax$ for the DustEM model in the $r$ and $g$ bands  is shown in Figure \ref{fig:polanglecomparison}.  

The case $\thetamax = 0$ requires no integration over the sphere. There is a single incident direction and a single scattering angle, $\theta = 90^\circ + b$. This corresponds to the backscattering lobe of \pha\ and \rinten\ is simply \pha\ at that $\theta$ ($130^\circ$ for Spider) and so less than 1.  The value of $p$ is simply $|\polpha/\pha|$ at that $\theta$. The scattering plane is equal to the meridional plane, and so the polarization arises simply from \polpha, which for the $r$ band is slightly negative according to Figure \ref{fig:polphasefxns}, top panel, and so $\psi = 90^\circ$. However, for the $g$ band, the polarization arises from the positive range of \polpha, and so $\psi$ flips to $0^\circ$ or equivalently $180^\circ$.  Note that an IFN location at a higher $b$ would also increase the scattering angle and induce the flip.

For increasing $\thetamax$, more scattering angles and dihedral angles come into play.  These are the same for the two bands and all dust models.  For the Sandage model, the illumination is also the same and so what determines the shape of each curve is the dust model, for which the dependence of \pha\ and \polpha\ on $\theta$ is unique (Figures \ref{fig:phasefxns} and \ref{fig:polphasefxns}). For this dust model at $g$,  there is a lot of cancellation among contributions to $Q_{\mathrm{tot}}$ and $U_{\mathrm{tot}}$, causing a rapid change in $\psi$ along with a minimum in $p$, before settling down to a behavior with $\thetamax$ like seen in the $r$ band.

For $\thetamax$ approaching $90^\circ$, the illumination is from a more and more extended slab filling the entire hemisphere. Still, all scattering angles are greater than $90^\circ$ and so \rinten\ is less than unity. We have computed the summary parameters only to $\thetamax = 89^\circ$ because of the rapid increase of the illumination that cannot be resolved for the finite $\nside = 256$ of the HEALpix representation.

Figure \ref{fig:polanglecomparison} also shows the corresponding results for the USIE illumination. As expected, they are identical to the Sandage results at $\thetamax = 0^\circ$. Furthermore, the values of $\tau_\infty$ are not so large as to produce a radical qualitative difference in the relative illumination with latitude at intermediate $\thetamax$, and so the net polarization is not much different than that produced by the Sandage model. However, at large $\thetamax$, the internal extinction limits the excessive growth of the illumination and the results depart from those of the Sandage model. Even at $\thetamax = 90^\circ$, \rinten, $\psi$, and $p$ can be calculated (e.g., the DustEM entries in Table \ref{tab:resultsu}).

\begin{deluxetable*}{lcc|ccc|ccc}
\tablecaption{Summary scattered light parameters for various dust models for the Spider IFN location ($\ell = 135^\circ, b = 40^\circ$)
\label{tab:resultsu}}
\tablewidth{0pt}
\tablehead{
\colhead{Model} & \colhead{$\varpi$} & \colhead{\gval} & \colhead{\rinten} & \colhead{$\psi$} & \colhead{$p$ [\%]} & \colhead{\rinten} & \colhead{$\psi$} & \colhead{$p$ [\%]}
}
\startdata
$r$ \\
DustEM & 0.6463 & 0.5453 & 0.5078 & 45\fdg55 & 9.403  & 0.6269 & 27\fdg26 & 3.996 \\
Astrodust & 0.7717 & 0.6463 & 0.4093 & 46\fdg66 & 9.234 &  0.5552 & 29\fdg13 & 3.600 \\
WD3.1 & 0.6412 & 0.5297 & 0.5379 & 46\fdg98 & 11.46 & 0.6453 & 29\fdg17 & 4.890\\
WD5.5 & 0.7476 & 0.5945 & 0.4620 & 52\fdg27 & 4.777 & 0.5960 & 38\fdg17 & 2.579 \\
\hline 
$g$ \\
DustEM & 0.6576 & 0.5690 & 0.4646 & 47\fdg66 & 8.101  & 0.6316 & 30\fdg87 & 3.156  \\
Astrodust & 0.7738 & 0.6821 & 0.3472 &  49\fdg76 & 5.193 & 0.5504 &  34\fdg28 & 1.881 \\
WD3.1 & 0.6743 & 0.5759 & 0.4606 & 47\fdg40 & 7.871 & 0.6332 & 31\fdg03 & 3.145 \\
WD5.5 & 0.7500 & 0.6191 & 0.4037 & 51\fdg31 & 1.478 & 0.5923 & 37\fdg90 & 1.389 \\
\enddata
\tablecomments{
Columns 4 -- 6: location parameter $b = 40^\circ$ and USIE illumination for $\thetamax = 90^\circ$.
Columns 7 -- 9: location ($135^\circ, 40^\circ$) and height 0.2 kpc above the plane and the relevant FRaNKIE models described in Section \ref{sec:fraill}.
}
\end{deluxetable*}

\subsubsection{\frankie}\label{subsubsec:frankie}

Still for the DustEM models ($r$ and $g$), we specify the illumination to be a \frankie\ model. Now both $\ell$ and $b$ need to be specified, because the radiation field is not axisymmetric. A height needs to be specified as well to select among the available \frankie\ models described in Section \ref{sec:fraill}.  Focusing again on the Spider IFN, we adopt location ($135^\circ, 40^\circ$) and height $z = 0.2$ kpc above the plane.

The summary parameters are tabulated in the DustEM rows and columns 7 -- 9 in Table \ref{tab:resultsu}.
Compared to the results for the full hemisphere USIE illumination model ($\thetamax = 90^\circ$), shown in Figure \ref{fig:polanglecomparison} and in columns 4 -- 6 in Table \ref{tab:resultsu}, the value of \rinten\ is higher, $\psi$ is slightly lower, and $p$ is significantly reduced. Again, this shows the possibility of discriminating between different aISRFs.

\subsection{Results for various dust models with the same aISRF} \label{subsec:varydust}

For the same USIE illumination with $\thetamax = 90^\circ$, and the Spider IFN location ($b=40^\circ$), results for several different dust models are compared in columns 4 -- 6 of Table \ref{tab:resultsu}. This illustrates the additional discriminatory power of scattered light polarization observations in assessing dust models, adding to the constraints from joint modeling of the intensity of thermal dust emission and scattered light intensity described by \citet{zhang}.

Table \ref{tab:resultsu} shows, in columns 7 -- 9, that this dust model discrimination is also the case with the \frankie\ illumination used in Section \ref{subsubsec:frankie}.

\section{Results for DustEM for varying IFN location}  
\label{sec:varycirrus}

The intensity and polarization parameters are sensitive to the location of the IFN being observed in scattered light.

\begin{figure}
    \centering
    \includegraphics[scale=0.5]{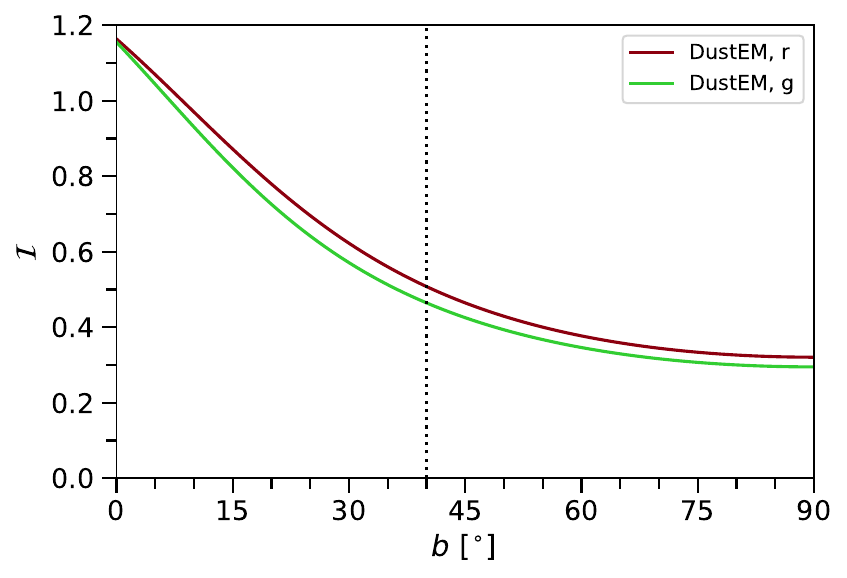}
    \includegraphics[scale = 0.5]{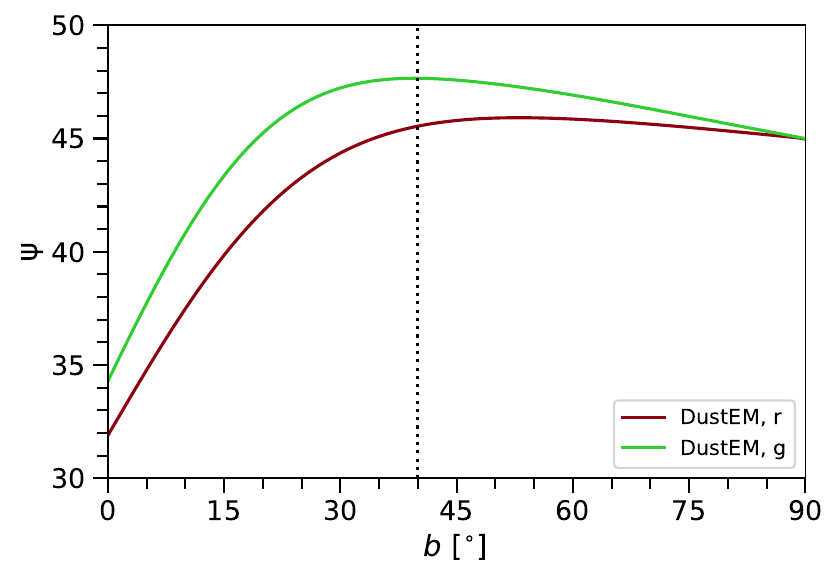}
    \includegraphics[scale = 0.5]{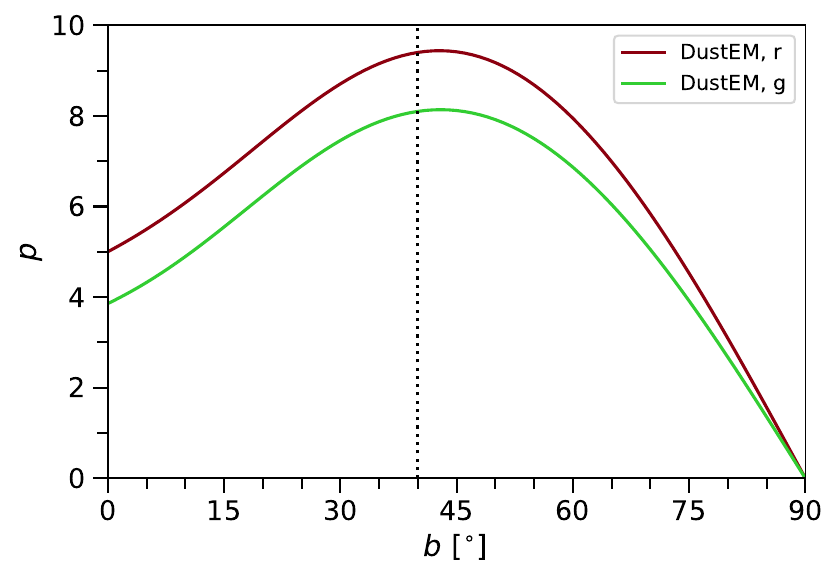}
    \caption{Effect of IFN location in Galactic latitude $b$ ($\ell = 135^\circ$) on summary scattered light parameters for USIE illumination ($\thetamax=90^\circ$) and DustEM grain model in the $r$ and $g$ bands.
    Top: relative intensity. Middle: polarization angle in degrees. Bottom: polarization fraction in percent.
    The vertical line corresponds to calculations for the Spider IFN in columns 4 -- 6 and DustEM rows in Table \ref{tab:resultsu}.
    }
    \label{fig:usieb}
\end{figure}

\subsection{Location as a function of $b$, for USIE illumination with $\thetamax=90^\circ$}
\label{subsubsec:sandageb}

Holding constant the dust model (DustEM) and the illumination (USIE with $\thetamax = 90^\circ$), we explored the sensitivity to $b$ that arises through \polpha\ (there is no dependence on Galactic longitude). This can be calculated over the full range $0^\circ$ to $90^\circ$. The results are shown in Figure \ref{fig:usieb}. The summary intensity and polarization parameters in the $r$ and $g$ bands follow similar trends, differing only because of slight differences in \pha\ (Figure \ref{fig:phasefxns}) and \polpha\ (Figure \ref{fig:polphasefxns}).

One might question the physical basis for the model at $b$ near $0^\circ$. It is the only case where \rinten\ is greater than unity because some radiation is scattered in the forward scattering lobe of \pha.  There \polpha\ becomes small, suppressing the resulting $p$. As $b$ increases, so do scattering angles $\theta$, and \rinten\ decreases while $p$ increases.

At $b$ near $90^{\circ}$, the IFN is almost directly above the slab with scattering angles in the range [$90^{\circ}$, $180^{\circ}$]. Cancellations among finite contributions from different illumination directions result in both $Q_{\mathrm{s,tot}}$ and $U_{\mathrm{s,tot}}$ near zero, so that $p$ approaches zero. However, the distribution of $Q_\mathrm{sp}$, while symmetrical about the direction $\ell$ of the IFN, is displaced along it. Furthermore, $\sin2\phi$ has the same sign in the hemisphere centered on that $\ell$, while $\cos2\phi$ does not, so that $|U_{\mathrm{s,tot}}|$ is orders of magnitude greater than $|Q_{\mathrm{s,tot}}|$. For this radiation field $\psi$ approaches $45^{\circ}$ (entirely net +U in Figure \ref{fig:stokes}).

\subsection{Location as a function of height, for \frankie\ illumination, $\ell=135^\circ$, and either constant $b=40^\circ$ or changing $b$}
\label{subsubsec:frankiebh}

\begin{deluxetable*}{lcccc|cccc} 
\tablecaption{Summary scattered light parameters for \frankie\ illumination at various heights $z$ above the Galactic plane, $\ell = 135^\circ$, and DustEM model} 
\label{tab:resultsf}
\tablewidth{0pt}
\tablehead{
\colhead{$z$ [kpc]} & \colhead{$b$} & \colhead{\rinten} & \colhead{$\psi$} & \colhead{$p$ [\%]} & \colhead{$b$} & \colhead{\rinten} & \colhead{$\psi$} & \colhead{$p$ [\%]}
}
\startdata
$r$ \\
0.2 & $40^\circ$ & 0.6269 & 27\fdg26 & 3.996  & $40^\circ$ & 0.6269 & 27\fdg26 & 3.996 \\
0.5 & $40^\circ$ & 0.5060 & 30\fdg52 & 4.751 & 64\fdg51 & 0.4212 & 05\fdg73 & 1.430 \\
1.0 & $40^\circ$ & 0.3862 & 30\fdg83 & 6.043 & 76\fdg59 & 0.3401 & 145\fdg4 &  4.238 \\
\hline 
$g$ \\
0.2 & $40^\circ$ & 0.6316 & 30\fdg87 & 3.156 & $40^\circ$ & 0.6316 & 30\fdg87 & 3.156 \\
0.5 & $40^\circ$ &  0.4874 & 37\fdg67 & 4.351 & 64\fdg51 & 0.3931 & 27\fdg27 & 1.262 \\
1.0 & $40^\circ$ & 0.3584 & 37\fdg83 & 5.863 & 76\fdg59 & 0.3131 & 150\fdg88 & 2.465 \enddata
\end{deluxetable*}

\frankie\ illumination at Galactocentric distance 8.5 kpc is available at discrete heights $z$ [kpc] above the Galactic plane, ranging from 0 to 30 kpc. Though it instructive to examine all of these, we take the models at $z = [0.2,0.5,1]$ kpc as most likely to be relevant for IFNs.

For one set of illustrative polarization calculations we hold $b$ constant, so that increasing $z$ corresponds to increasing distance from the Sun ($d = 0.31 (z/0.2)\sin(40)/\sin(b)$ kpc). We adopt ($\ell, b) = (135^\circ, 40^\circ)$ and the dust model DustEM so that we overlap the previous calculations in Section \ref{subsubsec:frankie} and Table \ref{tab:resultsu} (right). The results are tabulated in Table \ref{tab:resultsf}, columns 3 -- 5. With increasing height and distance, \rinten\ decreases while $p$ increases and there is a modest increase in $\psi$.  Because the scattering geometry remains the same ($\theta$, $\phi$, \pha, \polpha), these systematic changes originate in the greater anisotropy of the illumination with respect to Galactic latitude.

For another set of illustrative calculations, we can pair a corresponding $b$ with $z$ via $b = \arctan(\tan(40) z/0.2)$, which corresponds to IFNs whose distance from the Sun projected in the Galactic plane is held constant. We again fix $\ell = 135^\circ$ and tabulate the results in Table \ref{tab:resultsf}, columns 7 -- 9. Now there is a larger decrease of \rinten\ with height, as $b$ increases. There is a substantial change in $\psi$ accompanied by a minimum in $p$. These arise from the additional effects of scattering geometry. Generally, \rinten, $\phi$, and $p$ are diagnostic of the aISRF and scattering geometry, for a given dust model (in this case DustEM).

\subsection{Location as a function of $\ell$, for \frankie\ illumination at $z = 0.2$ kpc and $b= 40^\circ$}
\label{subsubsec:frankiel}

\begin{figure}
    \centering
    \includegraphics[scale=0.5]{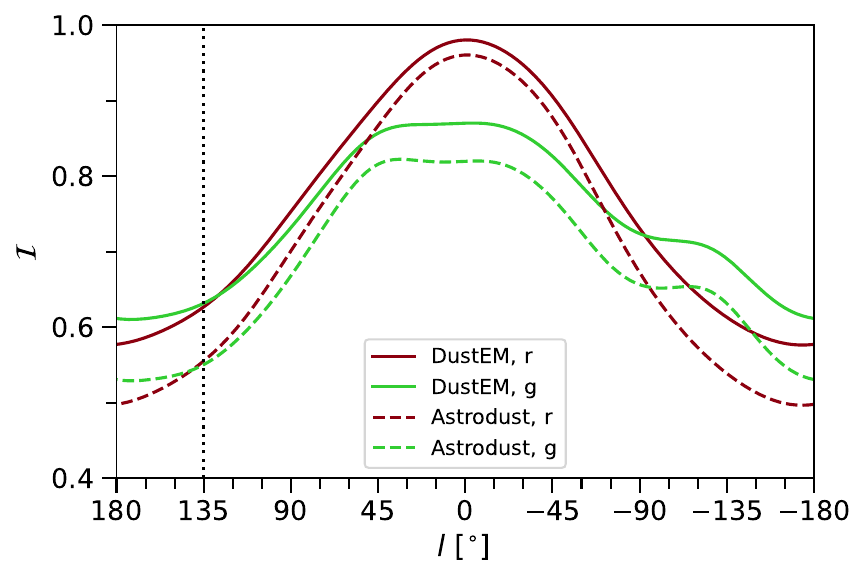}
    \includegraphics[scale = 0.5]{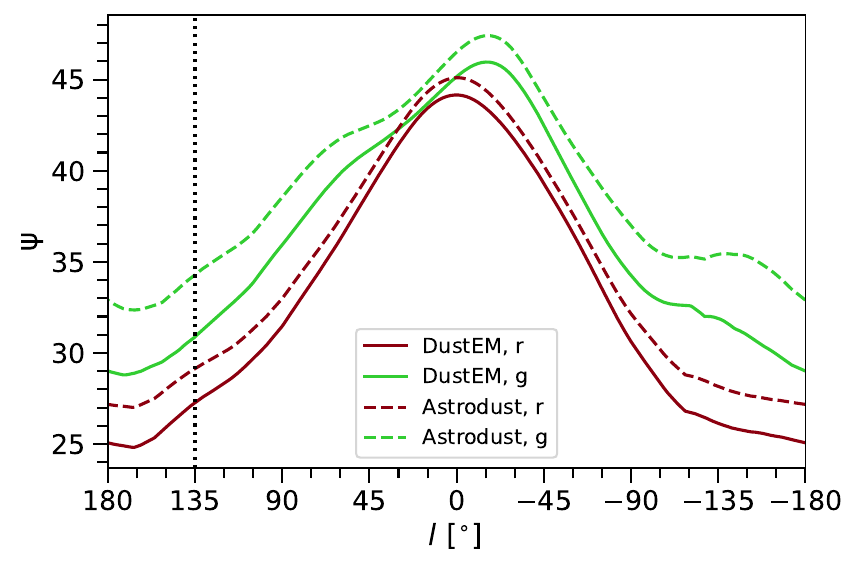}
    \includegraphics[scale = 0.5]{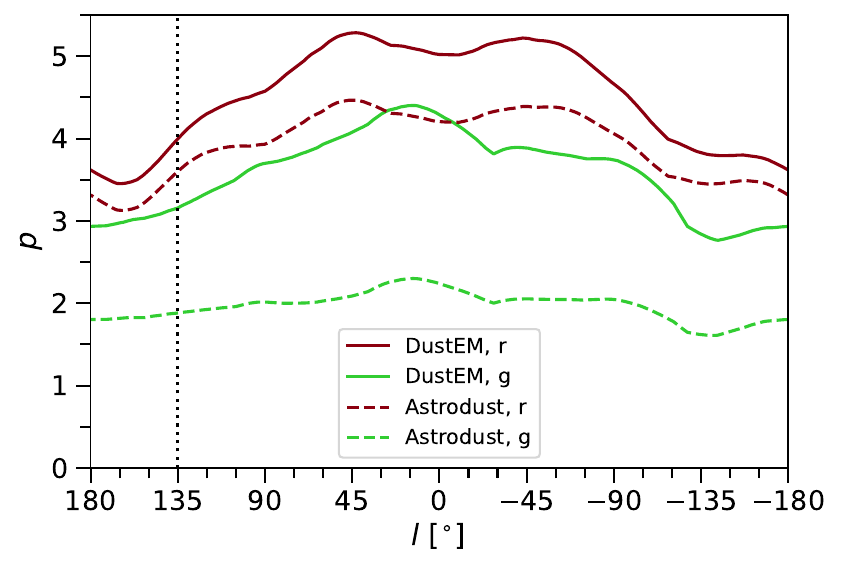}
    \caption{Effect of IFN location in Galactic longitude $\ell$ ($b= 40^\circ$) on summary scattered light parameters for \frankie\ illumination ($z = 0.2$ kpc) and two dust models in the $r$ and $g$ bands (DustEM and Astrodust). 
    Top: relative intensity. 
    Middle: polarization angle in degrees. 
    Bottom: polarization fraction in percent.
    The vertical line corresponds to calculations for the Spider IFN in columns 7 -- 9 and DustEM and Astrodust rows in Table \ref{tab:resultsu} and the $z =0.2$ kpc, $b = 40^\circ$ rows in Table \ref{tab:resultsf} for DustEM.
    }
    \label{fig:angle_position}
\end{figure}

Unlike the other illumination models, the \frankie\ models are anisotropic in longitude (as well as in latitude). This results in the variations of \rinten, $\psi$, and $p$ with $\ell$ shown in Figure \ref{fig:angle_position}, for $z= 0.2$ kpc and $b = 40^\circ$ and for DustEM (solid curves). The values at $\ell = 135^\circ$ were tabulated in Tables \ref{tab:resultsu} and \ref{tab:resultsf}.

In the $r$ band, the \rinten\ curves for both dust models  are not perfectly symmetrical about $\ell = 0^\circ$. There is an even greater asymmetry in $g$. The asymmetry reflects the fact that the \frankie\ illumination model is not symmetrical about $\ell = 0^\circ$. Nor in reality is the actual aISRF in the Milky Way expected to be symmetrical about $\ell = 0^\circ$, because of the distributions of stars (bar, spiral arms) and interstellar dust.

Recall that when the IFN is in a location relative to observer less favorable than average then \rinten $< 1$, because of the coupling of the $\theta$ dependence of the phase function $\pha$ and the illumination. The least favorable location is for IFNs toward the anticenter (around $\ell = \pm 180^\circ $), where backscattering is predominant. Locations in the direction of the Galactic bulge (around $\ell = 0^\circ $) are somewhat more favorable because of contributions from the forward scattering lobe. 

The $\psi$ and $p$ curves are also not symmetrical about $\ell = 0^\circ$. The polarization asymmetry is also greater in $g$. The polarization asymmetry now reflects the coupling of the more complex $\theta$ dependence of the polarization phase function $\polpha$ and the illumination asymmetry, with the latter having a dominant role.\footnote{This can be confirmed by pairing the \polpha\ for $r$ with the illumination for $g$, and vice versa.} 

\subsubsection{For the Astrodust model}
\label{sec:astroell}

We carried out the same calculations for the Astrodust model (dashed curves in Figure \ref{fig:angle_position}). The variation of $p$ with $\ell$ is smaller for Astrodust than for DustEM.  This is because of more cancellation of contributions from the positive and negative lobes of the Astrodust polarization phase function, \polpha\ (see Figure \ref{fig:polphasefxns}, top). As well, the similarity of the polarization parameter variations of Astrodust in the $r$ band and of DustEM in the $g$ band is related to the similarity of their corresponding \polpha\ in these respective bands.

\section{Discussion and Conclusions} \label{sec:discussion}

Considering optically thin IFNs, \citet{zhang} demonstrated that observations of thermal dust emission combined with any dust model and aISRF model could be used to predict the expected intensity of scattered light. Comparing this prediction with the observed intensity of scattered light tests combinations of dust models and aISRF models and when applied to the Spider IFN demonstrated that further refinement of the models was needed to get a closer match of predictions and observations. Furthermore, it was recognized that it would be difficult to pin down deficiencies in the dust models separately from those in the aISRF models.  Such joint modeling is related to the scattered light parameter \rinten, which we have shown above is also dependent on the location of the IFN in the Galaxy.  

We therefore undertook to predict the polarization properties of the scattered light as well, namely polarization angle $\psi$ and polarization fraction $p$, to assess whether they would be useful in providing complementary diagnostics of dust models and aISRF models to help break the degeneracies, including for IFNs in different locations. After developing the formalism for these calculations, we demonstrated systematically through many examples spanning different combinations of dust models, aISRFs, and Galactic locations that the potential for discrimination is there. 

However, it should be appreciated from these calculations that such observations are challenging, with optically thin IFNs being only a few percent of the brightness of the night sky. According to our results across many models, the polarization fraction is likely to be only a few percent and not more than 10\,\%.  This is much less than one might intuitively hope, based on familiarity with Rayleigh scattering in the daytime sky, where the magnitude of \polpha\ is much more favorable and the illumination is very anisotropic. This is a reminder that moonlight is to be avoided and that other sources of illumination of the night sky could induce an unwanted polarization contamination as well. Nevertheless, this contamination would be smooth across a field, unlike the intensity and polarized intensities of the scattered light that are modulated by the column density of dust in the IFN. The spatially different behavior would enable component separation.

Interstellar polarization produced by aligned dust grains in the foreground could contaminate the polarization as well.  It is unlikely to be oriented with the same $\psi$. If this contamination were not removed, and the wavelength dependence of $p$ of the scattered light were different than that of interstellar polarization, then the net $\psi$ would change (even more) with wavelength. For IFNs at intermediate to high Galactic latitudes, the interstellar polarization would likely be smaller than $p$ for the scattered light; it could be mapped for foreground stars whose distances are known from \Gaia\ \citep{gaia,gaiadr3} and its Stokes parameters subtracted.  

By comparison to simple diffuse Galactic light models, \citet{chellew2022} found evidence for an excess at 0.65 \micron\ that could arise from the ``extended red emission'' (ERE) phenomenon \citep{Witt2008}. The ERE is thought to be excited by near-UV photons and emitted isotropically, not subject to \pha. The ERE seems unlikely to be polarized. Its admixture in an IFN would therefore frustrate the simplest modeling in the $r$ band.

We have shown that location of the IFN is important. One promising IFN to contrast to the Spider is the Draco nebula discussed by \citet{zhang}. The Draco field is centered on ($91^\circ, 38^\circ$) and is at a distance of about 600 pc, thus at a height $z$ about 0.37 kpc above the Galactic plane. Being at a different Galactic location, the scattering geometry would provide a distinctly different coupling to the anisotropy of the aISRF. In Table \ref{tab:dracospider} we have previewed the magnitude of the differences that might be expected, using the DustEM model and \frankie\ illumination at various heights. The first row for each band reproduces the previous results for the Spider from Tables \ref{tab:resultsu} (right) and \ref{tab:resultsf}. The second row is for the Draco coordinates and the same \frankie\ illumination for 0.2 kpc, which is probably too low; because $b$ is similar to that for Spider, these results are close to those shown in Figure \ref{fig:angle_position}. The fourth row is for the Draco location but \frankie\ illumination for 0.5 kpc, which is probably too high. In the third row, we simply averaged the two \frankie\ files to provide an estimate for 0.35 kpc (note that the simple average of the values of \rinten\ for 0.2 and 0.5 kpc is very close to the tabulated value, the average for $\psi$ is only about $0\fdg045$ degrees lower, and the average for $p$ is higher by a factor less than 1.025). The Draco values are certainly distinguishable from those for the Spider.

\begin{deluxetable}{cccccc}
\tablecaption{Summary scattered light parameters for DustEM model and \frankie\ illumination for different heights for Spider and Draco locations}
\label{tab:dracospider}
\tablewidth{0pt}
\tablehead{
 \colhead{$z$ [kpc]} & \colhead{$\ell$} & \colhead{$b$} & \colhead{\rinten} & \colhead{$\psi$} & \colhead{$p$ [\%]}
}
\startdata
$r$\\
0.2  & 135 & 40 & 0.6269 & 27\fdg26 & 3.996  \\
0.2  & 91  & 38 & 0.7698 & 30\fdg90 & 4.571 \\
0.35 & 91  & 38 & 0.6871 & 31\fdg55 & 5.499 \\
0.5  & 91  & 38 & 0.6054 & 32\fdg11 & 6.670 \\
\hline
$g$ \\
0.2  & 135 & 40 & 0.6316 & 30\fdg87 & 3.156 \\
0.2  & 91  & 38 & 0.7491 & 35\fdg32 & 3.689 \\
0.35 & 91  & 38 & 0.6665 & 36\fdg42 & 4.351 \\
0.5  & 91  & 38 & 0.5828 & 37\fdg44 & 5.221 \\
\enddata
\end{deluxetable}

In conclusion, we have found that observations of \rinten, $\psi$, and $p$ of scattered light, along with the intensity of thermal dust emission, are a powerful combination with which to assess the viability of dust models and models of the aISRF, but they need to be carried out with care and precision. 

Looking at the various tables and figures, one can see that depending on the combinations of aISRF, dust model, and location either $r$ or $g$ could offer more discrimination.  Quite often $g$ might be preferred.  On the other hand, both the brightness and the polarization fraction can be lower at $g$, making observations more challenging; this would worsen to the blue. Moving further to the red than $r$, the brightness is again lower \citep{zhang}.   Overall, using a combination of both $r$ and $g$ seems optimal, rather than devoting observational resources to just one.

Both \Planck\ thermal dust polarization and interstellar polarization of starlight indicate that in reality dust particles are aspherical and aligned with respect to the ambient magnetic field \citep{planck2016-l11B} though the details in any particular IFN are presently difficult to prescribe. Though the nature of the asphericity is uncertain, the scattered light properties might be tackled in a next approximation with calculations for aligned spheroids, where \pha\ and \polpha\ would depend on the direction of the incident radiation with respect to the instantaneous orientation of the spheroid. There would be averaging for spinning/tumbling spheroids over a range of shapes and sizes and illumination by the all-sky aISRF and one might reasonably expect (though it has not been demonstrated) that the net results would be close to those obtained for spheres and that trends in results for the $r$ band relative to the $g$ band would be preserved. 

We can expect substantial improvement both in our understanding of how the aIRSF changes with position in the Galaxy, for example thanks to constraining data now available from \Gaia\ \citep{gaia,gaiadr3}, and in dust models. These will have a salutary impact on the testing previewed here, limiting the parameter space for model solutions consistent with the data.

\begin{acknowledgments}
We  acknowledge  support  from  the  Natural  Sciences and Engineering Research Council (NSERC) of Canada. SB acknowledges a Bell Internship through Mount Allison University and participation in the Summer Undergraduate Research Fellowship (SURF) program hosted by CITA at the University of Toronto. This research made use of the NASA Astrophysics Data System.
\end{acknowledgments}

\software{Some of the results in this paper have been calculated using AstroPy,\footnote{\url{http://www.astropy.org}} a community-developed core Python package for Astronomy \citep{astropy_2013, astropy_2018}, the HEALPix package and healpy \citep{gorski05, zonca19}, Matplotlib \citep{hunter_2007}, NumPy \citep{van_der_walt_2011}, and  SciPy \citep{SciPy-NMeth}.
}

\bibliography{paper_accepted}{}
\bibliographystyle{aasjournal}

\end{document}